\documentstyle[11pt]{article} 
\input epsf 

\def\msun{M_{\odot}} 
\def\mpl{m_{{\rm Pl}}}
\def\delh{\delta_{{\rm H}}}
\def\x{{\bf x}} 
\def\k{{\bf k}} 
 
\def\0{{\bf 0}} 
\def\1{{\bf 1}} 
\def\2{{\bf 2}}

\begin{document} 

\begin{center}
{\Large {\bf THE EARLY UNIVERSE\footnote{Preprint SUSSEX-AST 96/12-1, 
astro-ph/9612093 \\Lectures given at the Winter 
School `From Quantum Fluctuations to Cosmological Structures' in 
Casablanca, Morocco, December 1996. Please notify any errors to {\tt 
a.liddle@sussex.ac.uk}}}}\\
\vspace*{1cm}
{\large Andrew R.~Liddle} \\
{\em Astronomy Centre, University of Sussex, Falmer, Brighton BN1 
9QH,~~~U.~K.}

\vspace*{20pt}
{\bf Abstract}
\end{center}

\noindent
An introductory account is given of the modern understanding of the 
physics of the early Universe. Particular emphasis is placed on the paradigm 
of {\em cosmological inflation}, which postulates a period of accelerated 
expansion during the Universe's earliest stages. Inflation provides a 
possible origin for structure in the Universe, such as microwave background 
anisotropies, galaxies and galaxy clusters; these observed structures can 
therefore be used to test models of inflation. A brief account is given of 
other early Universe topics, namely baryogenesis, topological defects, dark 
matter candidates and primordial black holes.

\vspace*{0.5cm}
\begin{figure}[h]
\centering 
\leavevmode\epsfysize=10cm \epsfbox{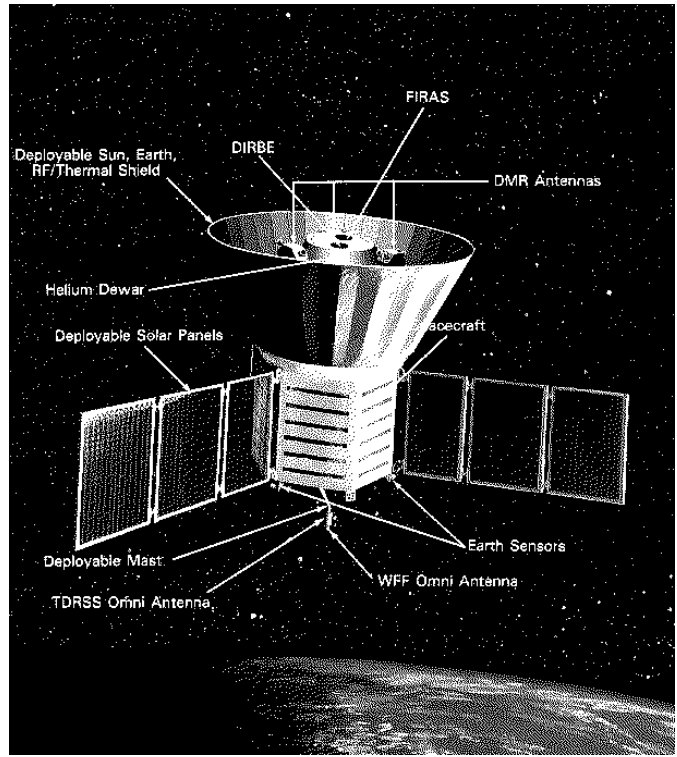}\\ 
\caption[cobe]{By providing the first measurement of irregularities in the 
microwave background radiation in 1992, the COBE satellite (artist's 
impression, courtesy NASA [reduced resolution in archive version]) 
revolutionized modern cosmology.} 
\end{figure} 
 
\newpage 
\tableofcontents 
 
\newpage
\section{Introduction} 
 
The early universe has been a major topic of research now for more than 
twenty years, and has come to encompass a wide range of different topics. 
The common theme is the introduction of ideas from particle physics, 
such as symmetry breaking and the unification of fundamental forces, into a 
cosmological setting. Conventionally, the `Early Universe' refers to those 
epochs during which the Universe was so hot and energetic that the 
appropriate physics is at an energy scale inaccessible even to the largest 
particle accelerators. The key tool therefore is speculative extrapolation 
of known physics into the realm of the unknown. Despite this, however, many 
of the key ideas can be understood without requiring any deep knowledge of 
particle physics, and in these lectures I shall aim to discuss the very 
early Universe from as astronomical a viewpoint as possible.

Although as a subject the early Universe has become quite a mature one, some 
of the most spectacular developments have been extremely recent, within the 
last few years. This is because for the first time it is becoming possible 
to probe the physics of the early Universe observationally, through the 
increasing range of observations of structure in the Universe. One of the 
most important realizations in this subject was that the standard cosmology, 
based on known physics, is incapable of providing a theory for the origin of 
structure in the Universe. Instead, any theory for that origin must lie in 
the early Universe. This has become the most important motivation for 
studying this subject, and has the important consequence that observations 
of structure in the Universe can be used to constrain models of the early 
Universe. Indeed, for the first time many proposals for the physics of the 
early Universe have been ruled out, a sure sign that the area is becoming a 
proper, falsifiable, science. The pivotal point in these new developments 
was the discovery in 1992 of anisotropies in the microwave background 
radiation, by the COsmic Background Explorer (COBE) satellite (shown in 
Figure 1) \cite{COBE1,COBE4}.

A central element of early Universe cosmology is the paradigm of {\em 
cosmological inflation}. This proposes a period of accelerated expansion in 
the Universe's distant past. While introduced to solve a set of largely 
conceptual problems concerning the initial conditions for the big bang 
theory, it was rapidly realized that it also provides a theory for the 
origin of structure in the Universe. I shall devote most of my time to 
inflation, especially since it, amongst all early Universe topics, is of the 
most direct relevance to the other lectures at this School. Only near the 
end will I make some discussion of some other research areas which lie 
within the early Universe heading.

The rapid recent developments in this area have rendered some of the 
literature rather obsolete in parts, but there are several good references. 
The classic early Universe textbook is the one of that name, by Kolb \& 
Turner \cite{KT}. This is the only textbook to cover the entire early 
Universe subject area. The book by Linde \cite{Linbook} concentrates on 
inflation, with a strong emphasis on particle physics aspects. Studies of 
structure in the Universe have been the topic of several recent books 
\cite{LSS}. A very nice review of inflation, written specifically with 
astronomers in mind but unfortunately pre-dating COBE, is that of Narlikar 
\& Padmanabhan \cite{NP}, and a more recent review of the relation between 
inflationary cosmology and structure in the universe was given by myself and 
Lyth \cite{LLrep}.

\section{An Early Universe Overview}

The standard hot big bang theory is an extremely successful one, passing 
some 
crucial observational tests, of which I'd highlight five.
\begin{itemize}
\item The expansion of the Universe.
\item The existence and spectrum of the cosmic microwave background 
radiation.
\item The abundances of light elements in the Universe (nucleosynthesis).
\item That the predicted age of the Universe is comparable to direct age 
measurements of objects within the Universe.
\item That {\em given} the irregularities seen in the microwave background 
by COBE, there exists a reasonable explanation for the development of 
structure in the Universe, through gravitational collapse.
\end{itemize}
In combination, these are extremely compelling. However, the standard hot 
big bang theory is limited to those epochs where the Universe is cool enough 
that the underlying physical processes are well established and understood 
through terrestrial experiment. It does not attempt to address the state of 
the Universe at earlier, hotter, times. Furthermore, the hot big bang theory 
leaves a range of crucial questions unanswered, for it turns out that it can 
successfully proceed only if the initial conditions are very carefully 
chosen. The assumption of early Universe studies is that the mysteries of 
the conditions under which the big bang theory operates may be explained 
through the physics occurring in its distant, unexplored past. If so, 
accurate observations of the present state of the Universe may highlight the 
types of process occurring during these early stages, and perhaps even shed 
light on the nature of physical laws at energies which it would be 
inconceivable to explore by other means.

The types of question that Early Universe Cosmology strives to answer are 
the following.
\begin{itemize}
\item What governs the global structure of the Universe?
\begin{itemize}
\item Why is the large-scale Universe so close to spatial flatness?
\item Why is the matter in the Universe so homogeneously (ie evenly) 
distributed on large scales?
\end{itemize}
\item What is the origin of structure in the universe (microwave background 
anisotropies, galaxies, galaxy clusters, etc)?
\item Why is there far more matter than antimatter in the universe?
\item What is the nature of the matter in the Universe? Is there any dark 
matter, and if so how much and what are its properties?
\item What are the consequences of exotic particle theories at high 
energies?
\begin{itemize}
\item Are topological defects (domain walls, cosmic strings, monopoles) 
produced in the early universe?
\item Are primordial black holes produced in the early Universe?
\item Do unusual particles such as axions exist?
\end{itemize}
\end{itemize}

\section{A Big Bang Reminder}

In this Section I'll provide a brief reminder of the standard hot big bang 
theory, and establish the notation I'll use throughout.

\subsection{Equations of motion}

The hot big bang theory is based on the {\em cosmological principle}, which 
states that the Universe should look the same to all observers. That tells 
us that the universe must be homogeneous and isotropic, which in turn tells 
us which metric must be used to describe it. It is the Robertson--Walker 
metric
\begin{equation}
ds^2 = -dt^2 + a^2(t) \left[ \frac{dr^2}{1-kr^2} + r^2 \left( d\theta^2
	+ \sin^2 \theta \, d\phi^2 \right) \right] \,.
\end{equation}
Here $t$ is the time variable, and $r$--$\theta$--$\phi$ are (polar) 
coordinates. The constant $k$ measures the spatial curvature, with $k$ 
negative, zero and positive corresponding to open, flat and closed universes 
respectively. If $k$ is zero or negative, then the range of $r$ is from zero 
to infinity and the universe is infinite, while if $k$ is positive then $r$ 
goes from zero to $1/\sqrt{k}$. Usually the coordinates are rescaled to make 
$k$ equal to $-1$, $0$ or $+1$. The quantity $a(t)$ is the scale-factor of 
the Universe, which measures how rapidly it is expanding. The form of $a(t)$ 
depends on the type of material within the Universe, as we'll see.

If no external forces are acting, then a particle at rest at a given set of 
coordinates $(r,\theta,\phi)$ will remain there. Such coordinates are said 
to be {\em comoving} with the expansion. One swaps between physical (ie 
actual) and comoving distances via
\begin{equation}
\mbox{physical distance} = a(t) \times \mbox{comoving distance} \,.
\end{equation}

The expansion of the Universe is governed by the properties of material 
within it. This can be specified\footnote{I follow standard cosmological 
practice of setting the fundamental constants $c$ and $\hbar$ equal to one. 
This makes the energy density and mass density interchangeable (since the 
former is $c^2$ times the latter). I shall also normally use the Planck mass 
$\mpl$ rather than the gravitational constant $G$; with the convention just 
mentioned they are related by $G \equiv \mpl^{-2}$.} by the energy density 
$\rho(t)$ and the pressure $p(t)$. These are often related by an equation of 
state, which gives $p$ as a function of $\rho$; the classic examples are
\begin{eqnarray}
p & = & \frac{\rho}{3} \quad \quad \mbox{Radiation} \,, \\
p & = & 0 \quad \quad \mbox{Non-relativistic matter} \,.
\end{eqnarray}
In general though there need not be a simple equation of state; for example 
there may be more than one type of material, such as a combination of 
radiation and non-relativistic matter, and certain types of material, such 
as a scalar field (a type of material crucial for modelling inflation), 
cannot be described by an equation of state at all.

The crucial equations describing the expansion of the Universe are
\begin{eqnarray}
H^2 = \frac{8\pi}{3 \mpl^2} \, \rho - \frac{k}{a^2} \quad \quad & &
	\mbox{Friedmann equation} \\
\dot{\rho} + 3(\rho+p) \frac{\dot{a}}{a}  = 0  \quad \quad & &
	\mbox{Fluid equation}
\end{eqnarray}
where overdots are time derivatives and $H = \dot{a}/a$ is the Hubble 
parameter. The terms in the fluid equation contributing to $\dot{\rho}$ have 
a simple interpretation; the term $3H\rho$ is the reduction in density due 
to the increase in volume, and the term $3Hp$ is the reduction in energy 
caused by the thermodynamic work done by the pressure when this expansion 
occurs.

These can also be combined to form a new equation
\begin{equation}
\frac{\ddot{a}}{a} = - \frac{4\pi}{3\mpl^2} \left( \rho + 3p \right) \quad
	\quad \mbox{Acceleration equation} 
\end{equation}
in which $k$ does not appear explicitly.

The spatial geometry is flat if $k = 0$. For a given $H$, this requires that 
the density equals the critical density
\begin{equation}
\rho_{{\rm c}}(t) = \frac{3 \mpl^2 H^2}{8\pi} \,.
\end{equation}
Densities are often measured as fractions of $\rho_{{\rm c}}$:
\begin{equation}
\Omega(t) \equiv \frac{\rho}{\rho_{{\rm c}}} \,.
\end{equation}

The present value of the Hubble parameter is still not that well known, and 
is 
normally parametrized as
\begin{equation}
H_0 = 100 h \; {\rm km \, s}^{-1} \, {\rm Mpc}^{-1} = \frac{h}{3000} 
	\, {\rm Mpc}^{-1} \,,
\end{equation}
where $h$ is normally assumed to lie in the range $0.5 \leq h \leq 0.8$. The 
present critical density is
\begin{equation}
\rho_{{\rm c}}(t_0) = 1.88 \, h^2 \times 10^{-29} \, {\rm g \, cm}^{-3} =
	2.77 \, h^{-1} \times 10^{11} \, \msun/(h^{-1} {\rm Mpc})^3 \,.
\end{equation}

The simplest solutions to these equations arise when a simple equation of 
state is chosen
\begin{eqnarray}
&\mbox{Matter Domination~~} p = 0 : & \rho \propto a^{-3} 
	\quad \quad a(t) \propto t^{2/3} \\
&\mbox{Radiation Domination~~} p = \rho/3 : & \rho \propto a^{-4} 
	\quad \quad a(t) \propto t^{1/2} \\
&\mbox{Cosmological Constant~~} p = - \rho : & \rho = \mbox{constant}
	\quad a(t) \propto \exp (Ht) 
\end{eqnarray}

\subsection{Characteristic scales}

The big bang universe has two characteristic scales
\begin{itemize}
\item The Hubble time/length $H^{-1}$.
\item The curvature scale $a|k|^{-1/2}$.
\end{itemize}
The first of these gives the characteristic timescale of evolution of 
$a(t)$, and the second gives the distance up to which space can be taken as 
flat. As written above they are both physical scales; to obtain the 
corresponding 
comoving scale one should divide by $a(t)$. The ratio of these scales 
actually gives a measure of $\Omega$; from the Friedmann equation we find
\begin{equation}
\sqrt{|\Omega -1|} = \frac{H^{-1}}{a |k|^{-1/2}} \,.
\end{equation}

A crucial property of the big bang universe is that it possesses {\em 
horizons}; even light can only travel a finite distance since the start of 
the universe $t_*$, given by
\begin{equation}
d_{{\rm H}}(t) = a(t) \int_{t_*}^{t} \frac{dt}{a(t)} \,.
\end{equation}
For example, matter domination gives $d_{{\rm H}}(t) = 3t = 2H^{-1}$. 
In a big bang universe, $d_{{\rm H}}(t_0)$ is a good approximation to the 
distance to the surface of last scattering, since $t_0 \gg t_{{\rm 
decoupling}}$.

\section{Problems with the Big Bang}

In this Section I shall quickly review the original motivation for the 
inflationary cosmology. These problems were largely one of initial 
conditions. While historically these problems were very important, they are 
now somewhat marginalized as focus is instead concentrated on inflation as a 
theory for the origin of cosmic structure. 

\subsection{The flatness problem}

The Friedmann equation can be written in the form
\begin{equation}
|\Omega-1| = \frac{|k|}{a^2 H^2} \,.
\end{equation}
During standard big bang evolution, $a^2 H^2$ is decreasing, and so $\Omega$ 
moves away from one, eg
\begin{eqnarray}
& \mbox{Matter domination:} & |\Omega-1| \propto t^{2/3} \\
& \mbox{Radiation domination:} & |\Omega-1| \propto t
\end{eqnarray}
where the solutions apply provided $\Omega$ is close to one. So $\Omega = 1$ 
is an {\em unstable} critical point. Since we know that today $\Omega$ is 
certainly within an order of magnitude of one, it must have been much closer 
in the past, eg
\begin{eqnarray}
& \mbox{nucleosynthesis ($t \sim 1 \, {\rm sec}$)} : & |\Omega-1| < 
	{\cal O}(10^{-16}) \\
& \mbox{electro-weak scale ($t \sim 10^{-11} \, {\rm sec}$)} : & 
	|\Omega-1| < {\cal O}(10^{-27})
\end{eqnarray}
That is, hardly any choices of the initial density lead to a Universe like 
our own. Typically, the Universe will either swiftly recollapse, or will 
rapidly expand and cool below 3K within its first second of existence.

\subsection{The horizon problem}

Microwave photons emitted from opposite sides of the sky appear to be in 
thermal equilibrium at almost the same temperature. Yet there was no time 
for those regions to interact before the photons were emitted, because of 
the finite horizon size.
\begin{equation}
\int_{t_*}^{t_{{\rm dec}}} \frac{dt}{a(t)} \ll 
	\int^{t_0}_{t_{{\rm dec}}} \frac{dt}{a(t)} \,. 
\end{equation}
In fact, any regions separated by more than about 2 degrees would be 
causally separated at decoupling in the hot big bang theory.

\subsection{The monopole problem (and other relics)}

Modern particle theories predict a variety of `unwanted relics', which would 
violate observations. These include
\begin{itemize}
\item Magnetic monopoles.
\item Domain walls.
\item Supersymmetric particles.
\item `Moduli' fields associated with superstrings.
\end{itemize}
Typically, the problem is that these are expected to be created very early 
in the Universe's history, during the radiation era. But because they are 
diluted by the expansion more slowly than radiation (eg $a^{-3}$ instead of 
$a^{-4}$) it is very easy for them to become the dominant material in the 
universe, in contradiction to observations. One has to dispose of them 
without harming the conventional matter in the universe.

\subsection{Homogeneity and isotropy}

This discussion is a variant on the horizon problem discussion given 
previously. The COBE satellite sees irregularities on all accessible angular 
scales, from a few degrees upwards. In the simplest cosmological models, 
where these irregularities are intrinsic to the last scattering surface, the 
perturbations are on too large a scale to have been created between the big 
bang and the time of decoupling, because the horizon size at decoupling 
subtends only a degree or so. Hence these perturbations must have been part 
of the initial conditions.\footnote{Note though that it is not yet known for 
definite that there are large-angle perturbations intrinsic to the last 
scattering surface. For example, in a topological defect model such as 
cosmic strings, such perturbations could be generated as the microwave 
photons propagate towards us.}

If this is the case, then the hot big bang theory does not allow a 
predictive theory for the origin of structure. While there is no reason why 
it is required to give a predictive theory, this would be a major setback  
and disappointment for the study of structure formation in the Universe.

\section{The Idea of Inflation}

Seen with many years of hindsight, the idea of inflation is actually rather 
obvious. Take for example the Friedmann equation as used to analyze the 
flatness problem
\begin{equation}
|\Omega -1 | = \frac{|k|}{a^2 H^2} \,.
\end{equation}
The problem with the hot big bang model is that $aH$ always decreases, and 
so $\Omega$ is repelled away from one.

Reverse this! Define inflation to be any epoch where $\ddot{a} > 0$, an 
accelerated expansion. We can rewrite this in several different ways
\begin{eqnarray}
\mbox{INFLATION} \quad & \Longleftrightarrow \quad & \ddot{a} > 0 \\
  & \Longleftrightarrow \quad & \frac{d(H^{-1}/a)}{dt} < 0 \\
  & \Longleftrightarrow \quad & p < - \frac{\rho}{3}
\end{eqnarray}
The middle definition is my favourite, because it has the most direct 
geometrical interpretation. It says that the Hubble length, as measured in 
comoving coordinates, {\em decreases} during inflation. At any other time, 
the comoving Hubble length increases. This is the key property of inflation; 
although typically the expansion of the Universe is very rapid, the crucial 
characteristic scale of the Universe is actually becoming smaller, when 
measured relative to that expansion.

Since the successes of the hot big bang theory rely on the Universe having a 
conventional (non-inflationary) evolution, we cannot permit this 
inflationary period to go on forever --- it must come to an end early enough 
that the big bang successes are not threatened. Normally, then, inflation is 
viewed as a phenomenon of the very early universe, which comes to an end and 
is followed by the conventional behaviour. Inflation does not replace the 
hot big bang theory; it is a bolt-on accessory attached at early times to 
improve the performance of the theory.

\subsection{The flatness problem}

\begin{figure}[t]
\centering 
\leavevmode\epsfysize=8.7cm \epsfbox{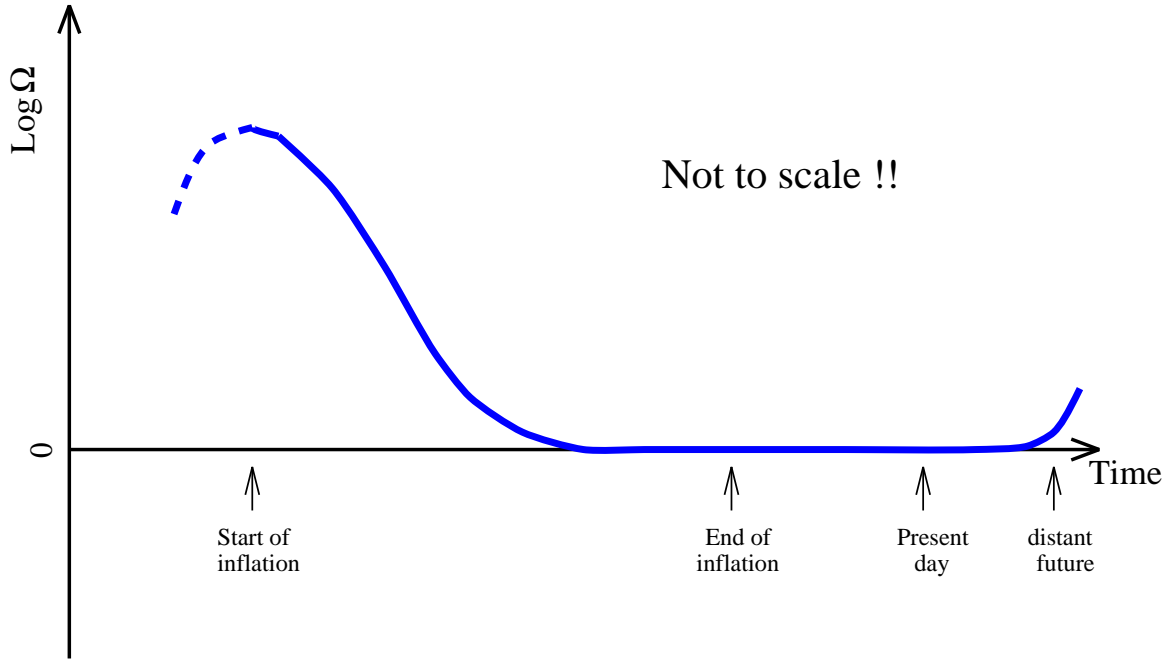}\\ 
\caption[flatness]{\label{flatness} A possible evolution of $\Omega$. There 
may or may not be evolution before inflation, shown by the dotted line. 
During inflation $\Omega$ is forced dramatically towards one, and remains 
there right up to the present. Only in the extremely distant future will it 
begin to evolve away from one again.} 
\end{figure}

We can now, more or less by definition, solve the flatness problem. From its 
definition (eg the middle condition above), inflation is precisely the 
condition that $\Omega$ is forced towards one rather than away from it. As 
we shall see, this typically happens very rapidly. All we need is to have 
enough inflation that $\Omega$ is moved so close to one during the 
inflationary epoch that it stays very close to one right to the present, 
despite being repelled from one for all the post-inflationary period. 
Obtaining sufficient inflation to perform this task is actually fairly easy.
A schematic illustration of this behaviour is shown in Figure 
\ref{flatness}.

\subsection{Relic abundances}

The rapid expansion of the inflationary stage rapidly dilutes the unwanted 
relic particles, because the energy density during inflation falls off more 
slowly (as $a^{-2}$ or slower) than the relic particle density. Very quickly 
their density becomes negligible.

This resolution can only work if, after inflation, the energy density of the 
Universe can be turned into conventional matter without recreating the 
unwanted relics. This can be achieved by ensuring that during the 
conversion, known as {\em reheating}, the temperature never gets hot enough 
again to allow their thermal recreation. Then reheating can generate solely 
the things which we want. Such successful reheating allows us to get back 
into the hot big bang universe, recovering all its later successes such as 
nucleosynthesis and the microwave background.

\subsection{The horizon problem and homogeneity}

The inflationary expansion also solves the horizon problem. The basic 
strategy is to ensure that
\begin{equation}
\int_{t_*}^{t_{{\rm dec}}} \frac{dt}{a(t)} \gg \int_{t_{{\rm dec}}}^{t_0}
	\frac{dt}{a(t)} \,,
\end{equation}
so that light can travel much further before decoupling than it can 
afterwards. This cannot be done with standard evolution, but can be achieved 
by inflation.

\begin{figure}[t]
\centering 
\leavevmode\epsfysize=12cm \epsfbox{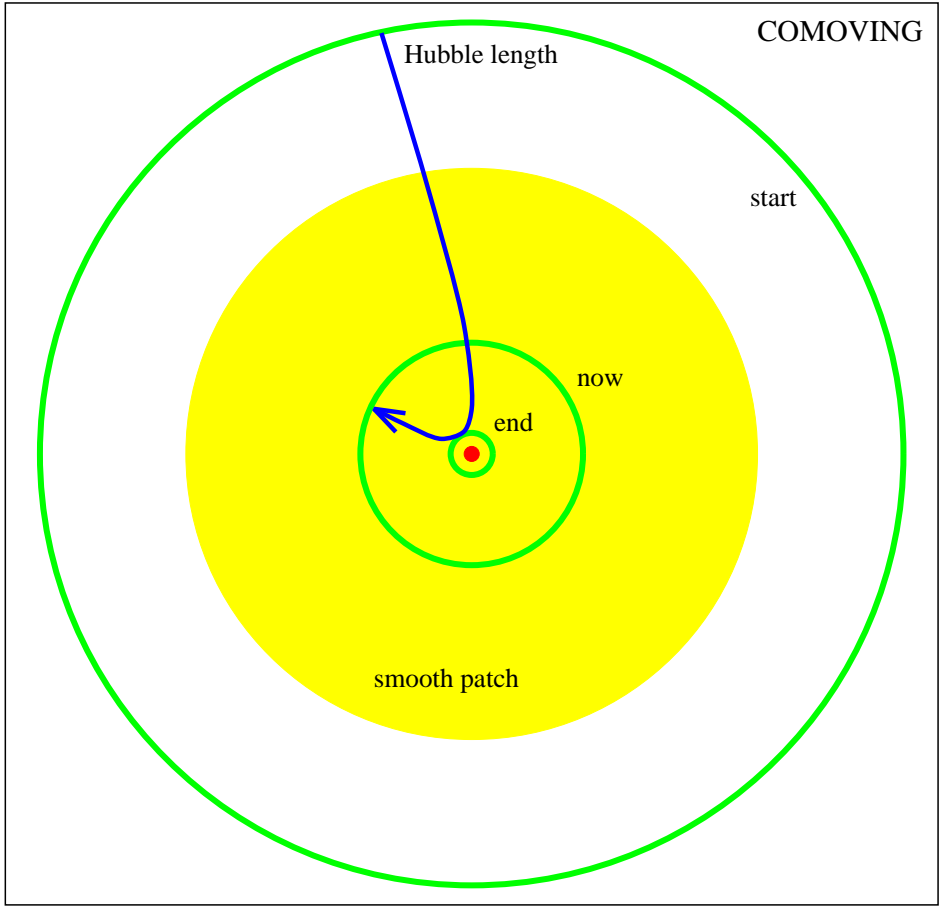}\\ 
\caption[horizon]{\label{horizon} Solving the horizon problem. Initially the 
Hubble length is large, and a smooth patch forms by causal interactions. 
Inflation then shrinks the Hubble length, and even the subsequent expansion 
again after inflation leaves the observable universe within the smoothed 
patch.} 
\end{figure}

An alternative way to view this is to remember that inflation corresponds to 
a decreasing comoving Hubble length. The Hubble length is ordinarily a good 
measure of how far things can travel in the universe; what this is telling 
us is that the region of the Universe we can see after (even long after) 
inflation is much smaller than the region which would have been visible 
before inflation started. Hence causal physics was perfectly capable of 
producing a large smooth thermalized region, encompassing a volume greatly 
in excess of our presently observable universe. In Figure \ref{horizon}, the 
outer circle indicates the initial Hubble length, encompassing the shaded 
smooth patch. Inflation shrinks this dramatically inwards towards the dot 
indicating our position, and then after inflation it increases while staying 
within the initial smooth patch.\footnote{Although this is a standard 
description, it isn't totally accurate. A more accurate argument is as 
follows \cite{LLrep}. At the beginning of inflation particles are 
distributed in a set of modes. This may be a thermal distribution or 
something else; whatever, since the energy density is finite there will be a 
shortest wavelength occupied mode, e.g. for a thermal distribution 
$\lambda_{{\rm max}} \sim 1/T$. Expressed in physical coordinates, once 
inflation has stretched all modes including this one to be much larger than 
the Hubble length, the Universe becomes homogeneous. In comoving 
coordinates, the equivalent picture is that the Hubble length shrinks in 
until it's much smaller than the shortest wavelength, and the universe, as 
before, appears homogeneous.}

Equally, causal processes would be capable of generating irregularities in 
the Universe on scales greatly exceeding our presently observable universe. 
I'll have much more to say about that soon.

\section{Modelling the Inflationary Expansion}

We have seen that a period of accelerated expansion --- inflation --- is 
sufficient to resolve a range of cosmological problems. But we need a 
plausible scenario for driving such an expansion if we are to be able to 
make proper calculations. This is provided by cosmological scalar fields.

\subsection{Scalar fields and their potentials}

In particle physics, a scalar field is used to represent spin zero 
particles. It transforms as a scalar (that is, it is unchanged) under 
coordinate transformations. In a homogeneous universe, the scalar field is a 
function of time alone.

In particle theories, scalar fields are a crucial ingredient for spontaneous 
symmetry breaking. The most famous example is the Higgs field which breaks 
the electro-weak symmetry, whose existence is hoped to be verified at the 
Large Hadron Collider at CERN when it commences experiments next millenium. 
Scalar fields are also expected to be associated with the breaking of other 
symmetries, such as those of Grand Unified Theories, supersymmetry etc.

\begin{itemize}
\item Any specific particle theory (eg GUTS, superstrings) contains scalar 
fields.
\item No fundamental scalar field has yet been observed.
\item In condensed matter systems (such as superconductors, superfluid 
helium etc) scalar fields are widely observed, associated with any phase 
transition. People working in that subject normally refer to the scalar 
fields as `order parameters'.
\end{itemize}

The starting point I'll use for our investigation is the expressions for the 
effective energy density and pressure of a homogeneous scalar field, which 
I'll call $\phi$. They are
\begin{eqnarray}
\label{effrho}
\rho_{\phi} & = & \frac{1}{2} \dot{\phi}^2 + V(\phi) \, \\
\label{effp}
p_{\phi} & = & \frac{1}{2} \dot{\phi}^2 - V(\phi) \,.
\end{eqnarray}
One can think of the first term in each as a kinetic energy, and the second 
as a potential energy. The potential energy $V(\phi)$ can be thought of as a 
form of `configurational' or `binding' energy; it measures how much internal 
energy is associated with a particular field value. Normally, like all 
systems, scalar fields try to minimize this energy; however, a crucial 
ingredient which allows inflation is that scalar fields are not always very 
efficient at reaching this minimum energy state.

Note in passing that a scalar field cannot in general be described by an 
equation of state; there is no unique value of $p$ that can be associated 
with a given $\rho$ as the energy density can be divided between potential 
and kinetic energy in different ways.

\begin{figure}[t]
\centering 
\leavevmode\epsfysize=7cm \epsfbox{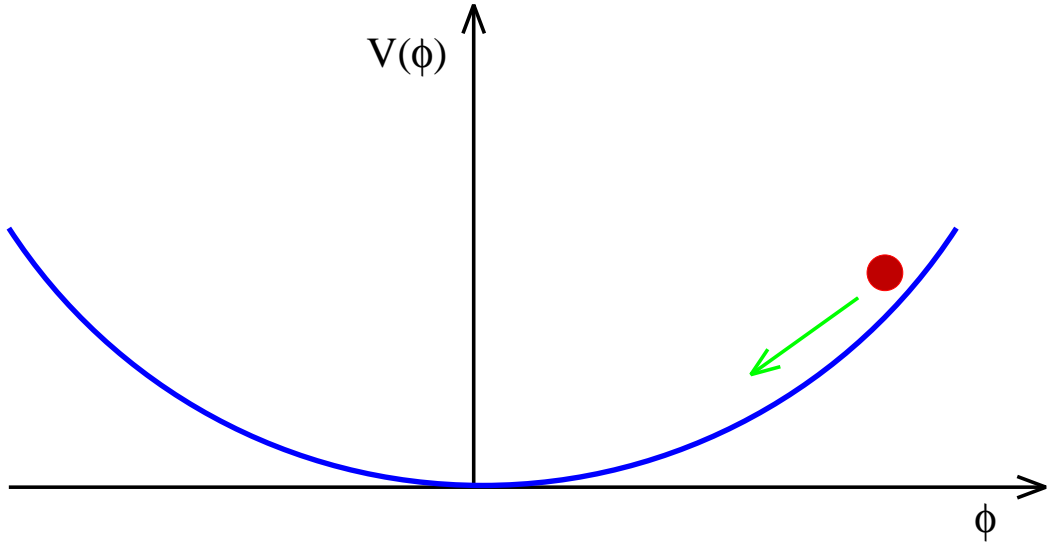}\\ 
\caption[scalpot]{\label{scalpot} A generic inflationary potential.} 
\end{figure}

In a given theory, there would be a specific form for the potential 
$V(\phi)$, at least up to some parameters which one could hope to measure 
(such as the effective mass and interaction strength of the scalar field). 
However, we are not presently in a position where there is a well 
established fundamental theory that one can use, so, in the absence of such 
a theory, inflation workers tend to regard $V(\phi)$ as a function to be 
chosen arbitrarily, with different choices corresponding to different models 
of inflation (of which there are many). Some example potentials are
\begin{eqnarray}
& V(\phi) = \lambda \left( \phi^2 - M^2 \right)^2 \quad \quad & 
	\mbox{Higgs potential} \\
& V(\phi) = \frac{1}{2} m^2 \phi^2 & \mbox{Massive scalar field}\\
& V(\phi) = \lambda \phi^4 & \mbox{Self-interacting scalar field}
\end{eqnarray}
The strength of this approach is that it seems possible to capture many of 
the crucial properties of inflation by looking at some simple potentials; 
one is looking for results which will still hold when more `realistic' 
potentials are chosen. Figure \ref{scalpot} shows such a generic potential, 
with the scalar field displaced from the minimum and trying to reach it.

\subsection{Equations of motion and solutions}

The equations for an expanding universe containing a homogeneous scalar 
field are easily obtained by substituting Eqs.~(\ref{effrho}) and 
(\ref{effp}) into the Friedmann and fluid equations, giving
\begin{eqnarray}
\label{scalfried}
H^2 & = & \frac{8\pi}{3 \mpl^2} \left[ V(\phi) + \frac{1}{2} \dot{\phi}^2
	\right] \,, \\
\label{scalwave}
\ddot{\phi} + 3 H \dot{\phi} & = & - V'(\phi) \,,
\end{eqnarray}
where prime indicates $d/d\phi$.

Since
\begin{eqnarray}
\ddot{a} > 0 & \Longleftrightarrow & p < - \frac{\rho}{3} \nonumber \\
  & \Longleftrightarrow & \dot{\phi}^2 < V(\phi)
\end{eqnarray}
we will have inflation whenever the potential energy dominates. This should 
be possible provided the potential is flat enough, as the scalar field would 
then be expected to roll slowly. The potential should also have a minimum in 
which inflation can end.

The standard strategy for solving these equations is the {\bf slow-roll 
approximation} (SRA); this assumes that a term can be neglected in each of 
the equations of motion to leave the simpler set
\begin{eqnarray}
H^2 & \simeq & \frac{8\pi}{3 \mpl^2} \, V \\
3 H \dot{\phi} & \simeq & -V'
\end{eqnarray}
If we define {\bf slow-roll parameters} \cite{LL}
\begin{equation}
\label{SR}
\epsilon(\phi) = \frac{\mpl^2}{16 \pi} \, \left( \frac{V'}{V}
	\right)^2 \quad ; \quad \eta(\phi) = \frac{\mpl^2}{8\pi}
	\, \frac{V''}{V} \,,
\end{equation}
where the first measures the slope of the potential and the second the 
curvature, then necessary conditions for the slow-roll approximation to hold 
are\footnote{Note that $\epsilon$ is positive by definition, whilst $\eta$ 
can have either sign.}
\begin{equation}
\epsilon \ll 1 \quad ; \quad |\eta| \ll 1 \,.
\end{equation}
Unfortunately, although these are necessary conditions for the slow-roll 
approximation to hold, they are not sufficient.
\begin{itemize}
\item The SRA reduces the order of the system of equations by one, and so 
its general solution contains one less initial condition. It works only 
because one can prove \cite{SB,LPB} that the solution to the full equations 
possesses an attractor property, eliminating the dependence on the extra 
parameter.
\item A more elaborate version of the SRA exists which is sufficient as well 
as necessary \cite{LPB}.
\item In Section \ref{infsr} I'll show, with some caveats, that if the 
slow-roll 
approximation is valid then one has inflation.
\end{itemize}

The amount of inflation is normally specified by the logarithm of the amount 
of expansion, {\em the number of e-foldings} $N$ given by
\begin{eqnarray}
N \equiv \ln \frac{a(t_{{\rm end}})}{a(t_{{\rm initial}})} & = &
	\int_{t_{{\rm i}}}^{t_{{\rm e}}} \, H \, dt \,, \\
 & \simeq & - \frac{8\pi}{\mpl^2} \int_{\phi_{{\rm i}}}^{\phi_{{\rm e}}}
 	\, \frac{V}{V'} \, d\phi \,,
\end{eqnarray}
where the final step uses the SRA. Notice that the amount of inflation 
between two scalar field values can be calculated without needing to solve 
the equations of motion, and also that it is unchanged if one multiplies 
$V(\phi)$ by a constant.

The minimum amount of inflation required to solve the various cosmological 
problems is about 70 $e$-foldings, i.e.~an expansion by a factor of 
$10^{30}$. Although this looks large, inflation is typically so rapid that 
most inflation models give much more.

\subsection{A worked example: polynomial chaotic inflation}

\label{quadratic}

The simplest inflation model \cite{Linbook} arises when one chooses a 
polynomial potential, such as that for a massive but otherwise 
non-interacting field, $V(\phi) = m^2 \phi^2/2$ where $m$ is the mass of the 
scalar field. With this potential, the slow-roll equations are
\begin{equation}
3H\dot{\phi} + m^2 \phi = 0 \quad ; \quad H^2 = \frac{4\pi m^2
	\phi^2}{3\mpl^2} \,,
\end{equation}
and the slow-roll parameters are
\begin{equation}
\epsilon = \eta = \frac{\mpl^2}{4 \pi \phi^2} \,.
\end{equation}
So inflation can proceed provided $|\phi| > \mpl/\sqrt{4\pi}$. The slow-roll 
solutions are
\begin{eqnarray}
\phi(t) & = & \phi_{{\rm i}} - \frac{m \, \mpl}{\sqrt{12\pi}} \, t \,, \\
a(t) & = & a_{{\rm i}} \exp \left[ \sqrt{\frac{4\pi}{3}} \, \frac{m}{\mpl}
	\, \left( \phi_{{\rm i}} t - \frac{m \, \mpl}{\sqrt{48\pi}} t^2
	\right) \right] \,,
\end{eqnarray}
(where $\phi = \phi_{{\rm i}}$ and $a = a_{{\rm i}}$ at $t = 0$) and the 
total amount of inflation is
\begin{equation}
\label{quadefold}
N_{{\rm tot}} = 2 \pi \, \frac{\phi_{{\rm i}}^2}{\mpl^2} - \frac{1}{2} \,.
\end{equation}
In order for classical physics to be valid we require $V \ll \mpl^4$, but it 
is still easy to get enough inflation provided $m$ is small enough. As we 
shall later see, $m$ is in fact required to be small from observational 
limits on the size of density perturbations produced.

\subsection{The relation between inflation and slow-roll}

\label{infsr}

The inflationary condition $\ddot{a} > 0$ is satisfied for a much wider 
range of behaviours than just (quasi-)exponential expansion. A classic 
example is power-law inflation $a \propto t^p$ for $p>1$, which is an exact 
solution for an exponential potential
\begin{equation}
V(\phi) = V_0 \exp \left[ - \sqrt{\frac{16 \pi}{p}} \, \frac{\phi}{\mpl}
	\right] \,.
\end{equation}

We can manipulate the condition for inflation as
\begin{eqnarray}
& &\frac{\ddot{a}}{a} = \dot{H} + H^2 > 0  \nonumber \\
\Longleftrightarrow & & -\frac{\dot{H}}{H^2} < 1 \nonumber \\
\stackrel{\sim}{\Longleftrightarrow} & & \frac{\mpl^2}{16\pi} \, \left( 
	\frac{V'}{V} \right)^2 < 1 \nonumber 
\end{eqnarray}
where the last manipulation uses the slow-roll approximation. The final 
condition is just the slow-roll condition $\epsilon < 1$, and hence
\[
\mbox{Slow-roll} \Longrightarrow \mbox{Inflation}
\]
However, the converse is not strictly true, since we had to use the 
SRA in the derivation. However, in practice
\begin{eqnarray}
\mbox{Inflation} & \stackrel{\sim}{\Longrightarrow} & \epsilon < 1 
	\nonumber \\
\mbox{Prolonged inflation} & \stackrel{\sim}{\Longrightarrow} & \eta < 1 
	\nonumber
\end{eqnarray}

\subsection{Reheating after inflation}

During inflation, all matter except the scalar field (usually called the 
inflaton) is redshifted to extremely low densities. {\bf Reheating} is the 
process whereby the inflaton's energy density is converted back into 
conventional matter after inflation, re-entering the standard big bang 
theory.

Once the slow-roll conditions break down, the scalar field begins to move 
rapidly on the Hubble timescale, and oscillates at the bottom of the 
potential. As it does so, it decays into conventional matter. The details of 
reheating (as long as we believe that it occurs!) are not important for our 
considerations here. I'll just note that recently there has been quite a 
dramatic change of view as to how reheating takes place. Traditional 
treatments (e.g.~Ref.~\cite{KT}) added a phenomenological decay term; this 
was constrained to be very small and hence reheating was viewed as being 
very inefficient. This allowed substantial redshifting to take place after 
the end of inflation and before the universe returned to thermal 
equilibrium; hence the reheat temperature would be lower, by several orders 
of magnitude, than suggested by the energy density at the end of inflation.

This picture is radically revised in work by Kofman, Linde \& Starobinsky 
\cite{KLS} (see also Ref.~\cite{preheat}), who suggest that the decay can 
undergo broad parametric resonance, with extremely efficient transfer of 
energy from the coherent oscillations of the inflaton field. This initial 
transfer has been dubbed {\em preheating}. With such an efficient start to 
the reheating process, it now appears possible that the reheating epoch may 
be very short indeed and hence that most of the energy density in the 
inflaton field at the end of inflation may be available for conversion into 
thermalized form. The full consequences of this change in viewpoint remain 
to be fully investigated.

\subsection{The range of inflation models}

Over the last ten years or so a great number of inflationary models have 
been devised, both with and without reference to specific underlying 
particle theories. Here I will discuss a very small subset of the models 
which have been introduced.

However, as we shall be discussing in the next Section, observations have 
great prospects for distinguishing between the different inflationary 
models. By far the best type of observation for this purpose appears to be 
high resolution satellite microwave background anisotropy observations, and 
we are fortunate that two proposals have recently been approved --- NASA has 
funded the {\em MAP} satellite \cite{MAP} for launch around 2000, and ESA 
has approved the hopefully soon to be renamed {\em COBRAS/SAMBA} satellite 
\cite{COBSAM} for launch a few years later. These satellites should offer 
very strong discrimination between the inflation models I shall now discuss. 
Indeed, it may even be possible to attempt a more challenging type of 
observation --- one which is independent of the particular inflationary 
model and hence begins to test the idea of inflation itself.

\subsubsection{Chaotic inflation models}

This is the standard type of inflation model \cite{Linbook}. The ingredients 
are
\begin{itemize}
\item A single scalar field, rolling in ...
\item A potential $V(\phi)$, which in some regions satisfies the slow-roll 
conditions, while also possessing a minimum with zero potential in which 
inflation is to end.
\item Initial conditions well up the potential, due to large fluctuations at 
the Planck era.
\end{itemize}
There are a large number of models of this type. Some are
\begin{quote}
\begin{tabbing}
Polynomial chaotic inflation \hspace*{0.3cm} \= $V(\phi) = 
	\frac{1}{2} m^2 \phi^2$ \hspace*{2cm} \= \\
 \> $V(\phi) = \lambda \phi^4$ \> \\
Power-law inflation \> $V(\phi) = V_0 \exp (\sqrt{\frac{16\pi}{p}} \,
	\frac{\phi}{\mpl})$ \> Exact solutions but ... \\
 \> \> No natural end to inflation. \\
`Natural' inflation \> $V(\phi) = V_0 [ 1 + \cos \frac{\phi}{f} ]$ \\
Intermediate inflation \> $V(\phi) \propto \phi^{-\beta}$ \> Also no 
	natural end.
\end{tabbing}
\end{quote}
Some of these actually do not satisfy the condition of a minimum in which 
inflation ends; they permit inflation to continue forever. However, we shall 
see power-law inflation arising in a more satisfactory context shortly.

\subsubsection{Multi-field theories}

\begin{figure}[tb]
\centering 
\leavevmode\epsfysize=8cm \epsfbox{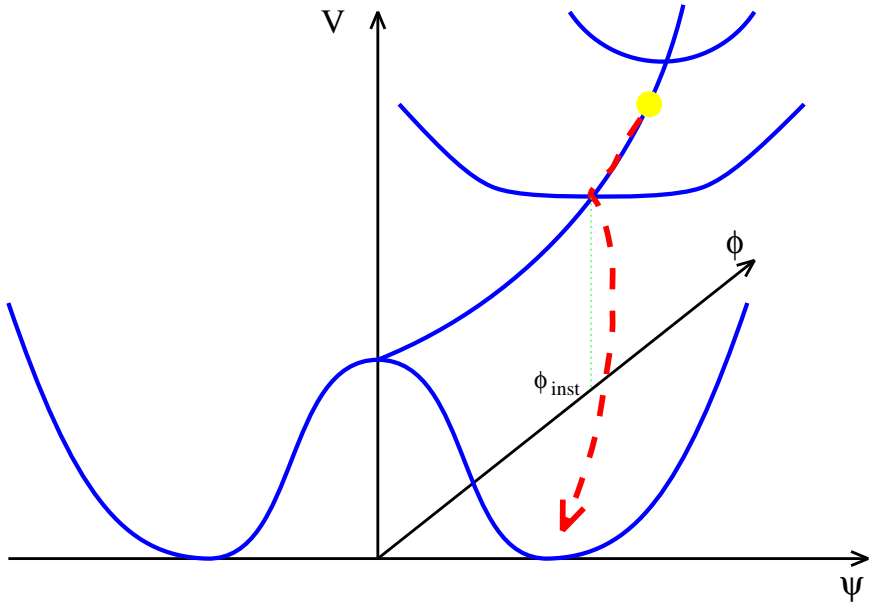}\\ 
\caption[hybridpot]{\label{hybridpot} The potential for the hybrid 
inflation model. The field rolls down the channel at $\psi = 0$ until it 
reaches the critical $\phi$ value, then falls off the side to the true 
minimum at $\phi = 0$ and $\psi = \pm M$.} 
\end{figure}

A recent trend in inflationary model building has been the exploration of 
models with more than one scalar field. The classic example is the hybrid 
inflation model \cite{hybrid}, which seems particularly promising for 
particle physics model building. It has a potential with two fields $\phi$ 
and $\psi$ of the form
\begin{equation}
V(\phi,\psi) = \frac{\lambda}{4} \left( \psi^2 - M^2 \right)^2 + 
	\frac{1}{2} m^2 \phi^2 + \frac{1}{2} \lambda' \phi^2 \psi^2 \,.
\end{equation}
which is illustrated in Figure \ref{hybridpot}. When $\phi^2$ is large, the 
minimum of the potential is at $\psi = 0$. The field rolls down this 
`channel' until it reaches $\phi_{{\rm inst}}^2 = \lambda M^2/\lambda'$, at 
which point $\psi = 0$ becomes unstable and the field rolls into one of the 
true minima at $\phi = 0$ and $\psi = \pm M$.

While in the `channel', which is where all the interesting behaviour takes 
place, this is just like a single field model with an effective potential 
for $\phi$ of the form
\begin{equation}
V_{{\rm eff}}(\phi) = \frac{\lambda}{4} M^4 + \frac{1}{2} m^2 \phi^2 \,.
\end{equation}
This is a fairly standard form, the unusual thing being the constant term, 
which would not normally be allowed as it would give a present-day 
cosmological constant. The most interesting regime is where that constant 
dominates, and it gives quite an unusual phenomenology. In particular, the 
energy density during inflation can be much lower than normal while still 
giving suitably large density perturbations, and secondly the field $\phi$ 
can be rolling extremely slowly which is of benefit to particle physics 
model building.

Within the more general class of two and multi-field inflation models, it is 
quite common for only one field to be dynamically important, as in the 
hybrid inflation model --- this effectively reduces the situation back to 
the single field case of the previous subsection. However, it may also be 
possible to have more than one important dynamical degree of freedom. In 
that case there is no attractor behaviour giving a unique route into the 
potential minimum, as in the single field case; for example, if the 
potential is of the form of an asymmetric bowl one could roll into the base 
down any direction. In that situation, the model loses its predictive power, 
because the late-time behaviour is not independent of the initial 
conditions.\footnote{Of course, there is no requirement that the `true' 
physical theory does have predictive power, but it would be unfortunate for 
us if it does not.}

\subsubsection{Beyond general relativity}

Rather than introduce an explicit scalar field to drive inflation, some 
theories modify the gravitational sector of the theory into something more 
complicated than general relativity \cite{W93}. Examples are
\begin{itemize}
\item Higher derivative gravity  ($R + R^2 + \cdots$).
\item Jordan--Brans--Dicke theory.
\item Scalar--tensor gravity.
\end{itemize}
The last two are theories where the gravitational constant may vary (indeed 
Jordan--Brans--Dicke theory is a special case of scalar--tensor gravity).

However, a clever trick, known as the {\em conformal transformation} 
\cite{conform}, allows such theories to be rewritten as general relativity 
plus one or more scalar fields with some potential. Often, only one of those 
fields is dynamical which returns us once more to the original chaotic 
inflation scenario!

The most famous example is extended inflation \cite{extinf}. In its original 
form, it transforms precisely into the power-law inflation model that we've 
already discussed, with the added bonus that it includes a proper method of 
ending inflation. Unfortunately though, this model is now ruled out by 
observations \cite{LL}! Indeed, models of inflation based on altering 
gravity are much more constrained than other types, since we know a lot 
about gravity and how well general relativity works \cite{W93}, and many 
models of this kind are very vulnerable to observations.

\subsubsection{Open inflation}

Recently, some interest has been given to an idea with a very long history 
indeed, which is a way of generating an open universe from 
inflation.\footnote{That is, a genuinely open universe with hyperbolic 
geometry and no cosmological constant.} Often in the past it has been 
declared that this is either impossible or contrived; however, it can be 
readily achieved in models with quantum tunnelling from a false vacuum (a 
metastable state) followed by a second inflationary stage \cite{open}. The 
tunnelling creates a bubble, and, incredibly, the region inside the 
expanding bubble looks just like an open universe, with the bubble wall 
corresponding to the initial (coordinate) singularity. These models are 
normally referred to as `open inflation' or `single-bubble' models.

These models are clearly very different from traditional inflation models, 
and may become the focus of extra attention should observational evidence 
for a low density universe continue to firm up. However I have no space to 
give them a proper discussion here and from now on will restrict discussion 
to the single-field chaotic inflation models.

\subsection{Recap}

The main points of this long Section are the following.
\begin{itemize}
\item Cosmological scalar fields, which were introduced long before 
inflation was thought of, provide a natural framework for inflation.
\item Despite a wide range of motivations, most inflationary models are 
dynamically equivalent to general relativity plus a single scalar field with 
some potential $V(\phi)$.
\item Within this framework, solutions describing inflation are easily 
found. Indeed, for many of the properties (amount of expansion, for 
example), we do not even need to solve the equations of motion.
\end{itemize}
With this information under our belts, we are now able to discuss the 
strongest motivation for the inflationary cosmology --- that it is able to 
provide an explanation for the origin of structure in the universe.

\section{Density Perturbations and Gravitational Waves}

In modern terms, by far the most important property of inflationary 
cosmology is that it produces spectra of both density perturbations and 
gravitational waves. The density perturbations may be responsible for the 
formation and clustering of galaxies, as well as creating anisotropies in 
the microwave background radiation. The gravitational waves do not affect 
the formation of galaxies, but as we shall see may contribute extra 
microwave anisotropies on the large angular scales sampled by the COBE 
satellite \cite{COBE4}.

Studies of large-scale structure typically make some assumption about the 
initial form of these spectra. Usually gravitational waves are assumed not 
to be present, and the density perturbations to take on a simple form such 
as the scale-invariant Harrison--Zel'dovich spectrum, or a scale-free 
power-law spectrum. It is clearly highly desirable to have a theory which 
predicts the forms of the spectra. There are presently two rival models 
which do this, {\em cosmological inflation} and {\em topological defects}. I 
will return to the topic of topological defects towards the end of these 
lectures.

\subsection{Production during inflation}

The ability of inflation to generate perturbations on large scales comes 
from the unusual behaviour of the Hubble length during inflation, namely 
that (by definition) the comoving Hubble length decreases. When we talk 
about large-scale structure, we are primarily interested in comoving scales, 
as to a first approximation everything is dragged along with the expansion. 
The qualitative behaviour of irregularities is governed by their scale in 
comparison to the characteristic scale of the Universe, the Hubble length. 

In the big bang universe the comoving Hubble length is always increasing, 
and so all scales are initially much larger than it, and hence unable to be 
affected by causal physics. Once they become smaller than the Hubble length, 
they remain so for all time. In the standard scenarios, COBE sees 
perturbations on large scales at a time when they were much bigger than the 
Hubble length, and hence no mechanism could have created them.

\begin{figure}
\centering 
\leavevmode\epsfysize=18cm \epsfbox{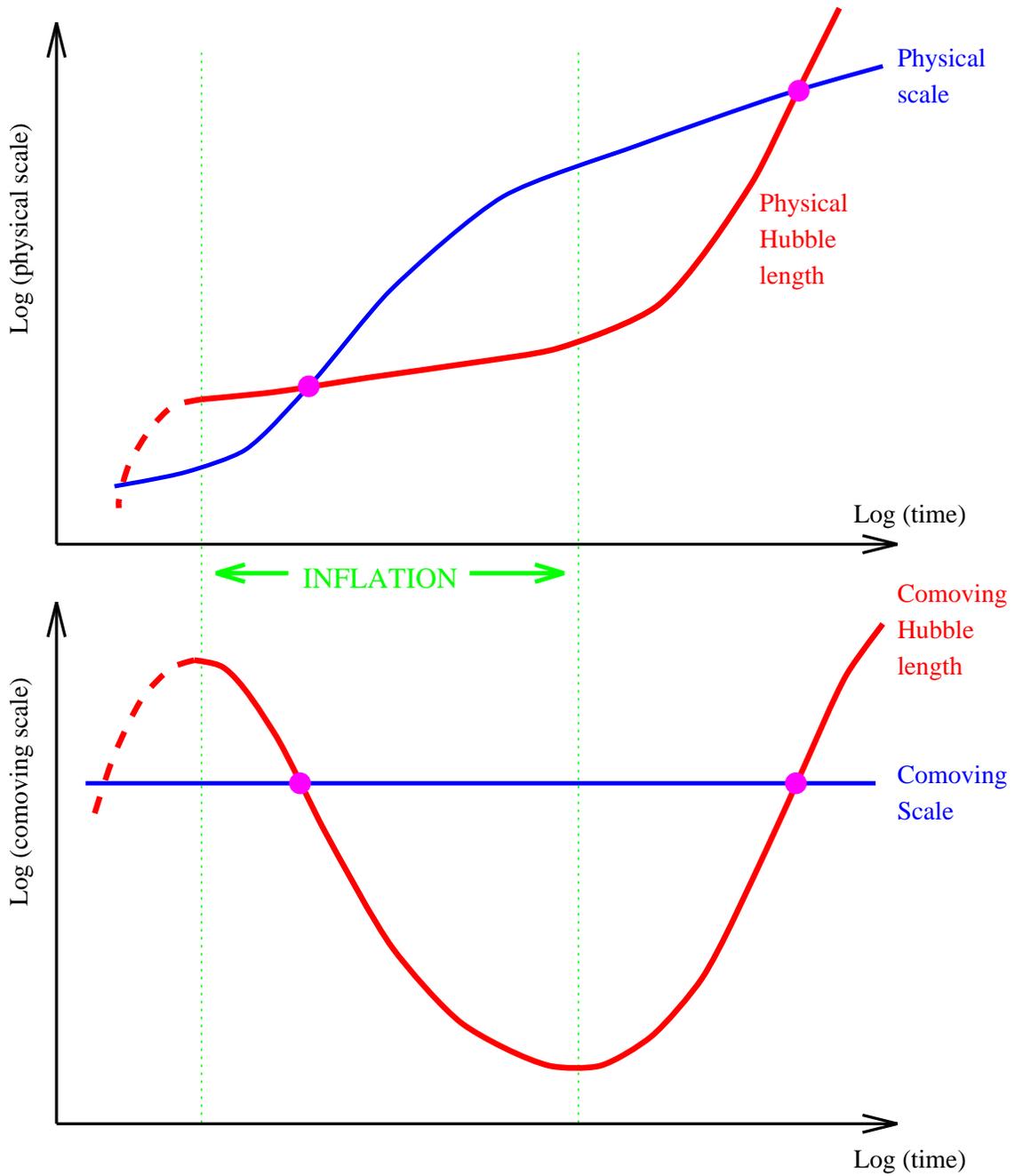}\\ 
\caption[scales]{\label{scales} The behaviour of a given comoving scale 
relative to the Hubble length, both during and after inflation, shown 
using physical coordinates (upper panel) and comoving ones (lower panel).} 
\end{figure} 

Inflation reverses this behaviour, as seen in Figure \ref{scales}. Now a 
given comoving scale has a more complicated history. Early on in inflation, 
the scale would be well inside the Hubble length, and hence causal physics 
can act, both to generate homogeneity to solve the horizon problem and to 
superimpose small perturbations. Some time before inflation ends, the scale 
crosses outside the Hubble radius (indicated by a circle in the lower panel 
of Figure \ref{scales}) and causal physics becomes ineffective. Any 
perturbations generated become imprinted, or, in the usual 
terminology, `frozen in'. Long after inflation is over, the scales cross 
inside the Hubble radius again. The perturbations we are interested in range 
from about the size of the present Hubble radius (i.e.~the size of the 
presently observable universe) down to a few orders of magnitude less. On 
the scale of Figure \ref{scales}, all interesting comoving scales lie 
extremely close together, and cross the Hubble radius during inflation very 
close together.

It's all very well to realize that the dynamics of inflation permits 
perturbations to be generated without violating causality, but we need a 
specific mechanism. That mechanism is quantum fluctuations. Inflation is 
trying as hard as it can to make the universe perfectly homogeneous, but it 
cannot defeat the Uncertainty Principle which ensures that there are always 
some irregularities left over. Through this limitation, it is possible for 
inflation to adequately solve the homogeneity problem and in addition 
leave enough irregularities behind to attempt to explain why the present 
universe is not completely homogeneous. 

The size of the irregularities depends on the energy scale at which 
inflation takes place. It is outside the scope of these lectures to describe 
in detail how this calculation is performed (see e.g.~Ref.~\cite{LLKCBA} for 
such a description); I'll just briefly outline the necessary steps and then 
quote the result, which we can go on to apply.
\begin{quote}
\begin{tabbing}
(a)~Perturb the scalar field \hspace*{3.5cm} \= $\phi = \phi(t) +
	\delta \! \phi(\x,t)$ \\
(b)~Expand in comoving wavenumbers \> $\delta \phi = \sum
	(\delta \! \phi)_{\k} e^{i\k.\x}$ \\
(c)~Linearized equation for classical evolution \> 
	$\ddot{\delta \! \phi}_{\k} +3H \dot{\delta \! \phi}_{\k} + 
	[ k^2/a^2 + V''] \delta \! \phi_{\k} =0 $ \\
(d)~Quantize theory   \\
(e)~Find solution with initial condition giving  \\
~~~~~flat space quantum theory ($k \gg aH$)  \\
(f)~Find asymptotic value for $k \ll aH$ \> $\langle | \delta \!
	\phi_{\k}|^2 \rangle = H^2/2k^3$ \\
(g)~Relate field perturbation to metric \> ${\cal R} = H \, \delta
	\! \phi/\dot{\phi}$ \\
~~~~~or curvature perturbation
\end{tabbing}
\end{quote}
Some important points are
\begin{itemize}
\item The details of this calculation are extremely similar to those used to 
calculate the Casimir effect (a quantum force between parallel plates), 
which has been tested in the laboratory.
\item The calculation itself is not controversial, though some aspects of 
its interpretation (in particular concerning the quantum to classical 
transition) are.
\item Exact analytic results are not known for general inflation models 
(though linear theory results for arbitrary models are easily calculated 
numerically). The results I'll be quoting will be lowest-order in the SRA, 
which is good enough for present observations.
\item Results are known to second-order in slow-roll for arbitrary inflaton 
potentials. Power-law inflation is the only standard model for which exact 
results are known. In some other cases, high accuracy approximations give 
better results (e.g.~small-angle approximation in natural or hybrid 
inflation).
\end{itemize}

The formulae for the amplitude of density perturbations, which I'll call 
$\delh(k)$, and the gravitational waves, $A_{{\rm G}}(k)$, are\footnote{The 
precise normalization of the spectra is arbitrary, as are the number of 
powers of $k$ included. I've made my favourite choice here (following 
\cite{LLrep,LLKCBA}), but whatever convention is used the normalization 
factor will disappear in any physical answer. For reference, the usual 
power spectrum $P(k)$ is proportional to $k \delh^2(k)$.}
\begin{eqnarray}
\label{delh}
\delh(k) & = & \left. \sqrt{\frac{512 \pi}{75}} \, \frac{V^{3/2}}{\mpl^3
	|V'|} \right|_{k = aH} \,, \\
\label{AG}
A_{{\rm G}}(k) & = & \left. \sqrt{\frac{32}{75}} \, \frac{V^{1/2}}{\mpl^2}
	\right|_{k = aH} \,.
\end{eqnarray}
Here $k$ is the comoving wavenumber; the perturbations are normally analyzed 
via a Fourier expansion into comoving modes. The right-hand sides of the 
above equations are to be evaluated at the time when $k = aH$ 
during inflation, which for a given $k$ corresponds to some particular value 
of $\phi$. We see that the amplitude of perturbations depends on the 
properties of the inflaton potential at the time the scale crossed the 
Hubble radius during inflation. The relevant number of $e$-foldings from the 
end of inflation is given by \cite{LLrep}
\begin{equation}
N \simeq 62 - \ln \frac{k}{a_0 H_0} + \mbox{numerical correction} \,,
\end{equation}
where `numerical correction' is a typically smallish (order one or a few) 
number which depends on the energy scale of inflation, the duration of 
reheating and so on. Normally it is a perfectly fine approximation to say 
that the scales of interest to us crossed outside the Hubble radius 60 
$e$-foldings before the end of inflation. Then the $e$-foldings formula
\begin{equation}
\label{efold2}
N \simeq - \frac{8\pi}{\mpl^2} \int_{\phi}^{\phi_{{\rm end}}} \, 
	\frac{V}{V'} \, d\phi \,,
\end{equation}
tells us the value of $\phi$ to be substituted into Eqs.~(\ref{delh}) and 
(\ref{AG}).

\subsection{A worked example}

The easiest way to see what is going on is to work through a specific 
example, the $m^2 \phi^2/2$ potential which we already saw in Section 
\ref{quadratic}. We'll see that we don't even have to solve the evolution 
equations to get our predictions.
\begin{enumerate}
\item Inflation ends when $\epsilon = 1$, so $\phi_{{\rm end}} \simeq 
\mpl/\sqrt{4\pi}$.
\item We're interested in 60 $e$-foldings before this, which from 
Eq.~(\ref{quadefold}) gives $\phi_{60} \simeq 3 \mpl$.
\item Substitute this in:
\[
\delh \simeq 12 \, \frac{m}{\mpl} \quad ; \quad A_{{\rm G}} \simeq 1.4
	\, \frac{m}{\mpl} 
\]
\item Reproducing the COBE result requires $\delh \simeq 2 \times 10^{-5}$ 
\cite{BLW} (provided $A_{{\rm G}} \ll \delh$), so we need $m \simeq 10^{-6} 
\mpl$.
\end{enumerate}

\subsection{Observational consequences}

Observations have moved on beyond us wanting to know the overall 
normalization of the potential. The interesting things are
\begin{enumerate}
\item The scale-dependence of the spectra.
\item The relative influence of the two spectra.
\end{enumerate}
These can be neatly summarized using the slow-roll parameters $\epsilon$ and 
$\eta$ we defined earlier \cite{LL}.

The standard approximation used to describe the spectra is the {\bf 
power-law approximation}, where we take
\begin{equation}
\delh^2(k) \propto k^{n-1} \quad ; \quad A_{{\rm G}}^2(k) 
	\propto k^{n_{{\rm G}}} \,,
\end{equation}
where the spectral indices $n$ and $n_{{\rm G}}$ are given by
\begin{equation}
n-1 = \frac{d \ln \delh^2}{d \ln k} \quad ; \quad n_{{\rm G}} =
	\frac{d \ln A_{{\rm G}}^2}{d \ln k} \,.
\end{equation}
The power-law approximation is usually valid because only a limited range of 
scales are observable, with the range $1$ Mpc to $10^4$ Mpc corresponding to 
$\Delta \ln k \simeq 9$.

The crucial equation we need is that relating $\phi$ values to when a scale 
$k$ crosses the Hubble radius, which from Eq.~(\ref{efold2}) is
\begin{equation}
\frac{d \ln k}{d \phi} = \frac{8\pi}{\mpl^2} \, \frac{V}{V'} \,.
\end{equation}
(since within the slow-roll approximation $k \simeq \exp N$). Direct 
differentiation then yields \cite{LL}
\begin{eqnarray}
n & = & 1 - 6\epsilon + 2 \eta \,, \\
n_{{\rm G}} & = & -2 \epsilon \,,
\end{eqnarray}
where now $\epsilon$ and $\eta$ are to be evaluated on the appropriate part 
of the potential.

Finally, we need a measure of the relevant importance of density 
perturbations and gravitational waves. The natural place to look is the 
microwave background; a detailed calculation which I cannot reproduce here 
(see e.g.~Ref.~\cite{LLrep}) gives
\begin{equation}
\label{relgrav}
R \equiv \frac{C_{\ell}^{{\rm GW}}}{C_{\ell}^{{\rm DP}}} \simeq \,
	4 \pi \epsilon \,.
\end{equation}
Here the $C_{\ell}$ are the contributions to the microwave multipoles, in 
the usual notation.\footnote{Namely, $\Delta T/T = \sum a_{\ell m} 
Y^{\ell}_{m}(\theta,\phi)$, $C_{\ell} = \langle | a_{\ell m}|^2 \rangle$.}

From these expressions we immediately see
\begin{itemize}
\item If and only if $\epsilon \ll 1$ and $|\eta| \ll 1$ do we get $n \simeq 
1$ and $R 
\simeq 0$.
\item Because the coefficient in Eq.~(\ref{relgrav}) is so large, 
gravitational waves can have a significant effect even if $\epsilon$ is 
quite a bit smaller than one.
\end{itemize}

Table \ref{pred} shows the predictions for a range of inflation models. Even 
the simplest inflation models can affect the large-scale structure modelling 
at a level comparable to the present observational accuracy. The predictions 
of the different models will be wildly different as far as future high 
accuracy observations are concerned.

{\small
\begin{table}
\begin{tabular}{|l|c|c|c|l|}
\hline
MODEL & POTENTIAL & $n$ & $R$& Note\\
\hline
\hline
Polynomial        & $\phi^2$ & 0.97 & 0.1 & \\
chaotic inflation & $\phi^4$ & 0.95 & 0.2 & \\
\hline
Power-law inflation & $\exp ( -\lambda \phi)$ & any $n<1$ & $2\pi(1-n)$ & 
Inc.~extended inflation \\
\hline
`Natural' inflation & $1 + \cos(\phi/f)$ & any $n<1$ & 0 & Basis for
	`tilted' models\\
\hline
Hybrid inflation (standard) & $1 + B\phi^2$ & 1 & 0 & Gives `simplest'
	spectra\\
Hybrid inflation (extreme)  & $1 + B\phi^2$ & $1 < n < 1.15$ & $\sim 0$
	& `Blue' spectra\\
\hline
\end{tabular}
\caption[pred]{\label{pred} The spectral index and gravitational wave 
contribution for a range of inflation models.}
\end{table}
}

Observations have some way to go before the power-law approximation becomes 
inadequate. Consequently ...
\begin{itemize}
\item Slow-roll inflation adds two, and only two, new parameters to 
large-scale structure.
\item Although $\epsilon$ and $\eta$ are the fundamental parameters, it is 
best to take them as $n$ and $R$.
\item Inflation models predict a wide range of values for these. Hence 
inflation makes no definite prediction for large-scale structure.
\item However, this means that large-scale structure observations, and 
especially microwave background observations, can strongly discriminate 
between inflationary models. When they are made, most existing inflation 
models will be ruled out.
\end{itemize}

\subsection{Tests of inflation}

The moral of the previous Section was that different inflation models lead 
to very different models of structure formation, spanning a wide range of 
possibilities. That means, for example, that a definite measure of say the 
spectral index $n$ would rule out most inflation models. But it would always 
be possible to find models which did give that value of $n$. Is there any 
way to try and test the idea of inflation, independently of the model 
chosen?

The answer, in principle, is yes. In the previous Section we introduced 
three observables (in addition to the overall normalization), namely $n$, 
$R$ and $n_{{\rm G}}$. However, they depend only on two fundamental 
parameters, namely $\epsilon$ and $\eta$ \cite{LL}. We can therefore 
eliminate $\epsilon$ and $\eta$ to obtain a relation between observables, 
the {\em consistency equation}
\begin{equation}
R = - 2 \pi n_{{\rm G}} \,.
\end{equation}
This relation has been much discussed in the literature \cite{recon,LLKCBA}. 
It is {\em independent} of the choice of inflationary model (though it does 
rely on the slow-roll and power-law approximations).

The idea of a consistency equation is in fact very general. The point is 
that we have obtained two continuous functions, $\delh(k)$ and $A_{{\rm 
G}}(k)$, from a single continuous function $V(\phi)$. This can only be 
possible if the functions $\delh(k)$ and $A_{{\rm G}}(k)$ are related, and 
the equation quoted above is the simplest manifestation of such a relation.

Vindication of the consistency equation would be a remarkably convincing 
test of the inflationary paradigm, as it would be highly unlikely that any 
other production mechanism could entangle the two spectra in the way 
inflation does. Unfortunately though, measuring $n_{{\rm G}}$ is a much more 
challenging observational task than measuring $n$ or $R$ and may be beyond 
even next generation observations. Indeed, this is a good point to remind 
the reader that even if inflation is right, only one model can be right and 
it is perfectly possible (and maybe even probable, see 
Ref.~\cite{lythnew}) that that model has a very low amplitude of 
gravitational waves and that they will never be detected.

\section{Further Early Universe Topics}

Most of my time has been spent discussing cosmological inflation and its 
consequences. It's time now to move on to a brief discussion of some of the 
other topics which fall under the umbrella of Early Universe Cosmology. Most 
of them are discussed at a high level of detail in the book by Kolb \& 
Turner \cite{KT}.

\subsection{Baryogenesis}

There is considerable observational evidence that the Universe is, by a vast 
majority, comprised of baryons rather than anti-baryons. Evidence within our 
solar system comes from the lack of annihilations experienced by any lunar 
or interplanetary probes, while from beyond the absence of antinucleons in 
cosmic rays, and of gamma rays produced in annihilations. The existence of 
horizons in the early Universe precludes the possibility of large-scale 
segregations needed to preserve a baryon-symmetric Universe.

In comparison to the number density of photons, the present number density 
of baryons is small indeed --- about one baryon per ten billion photons. 
However, if one tracks this asymmetry back to early times, assuming baryon 
number conservation, then at early times one expects the baryons to be in 
thermal equilibrium with the photons and hence with the same number density. 
At that time, the fractional baryon asymmetry would be very small indeed; 
for every ten billion photons there would be ten billion anti-baryons and 
ten 
billion and one baryons. Once the Universe cools enough, the ten billion 
anti-baryons annihilate with the baryons to leave the small excess.

Reminiscent of say the flatness problem, we could imagine that baryon number 
is perfectly conserved in the Universe, and that the excess of baryons over 
anti-baryons is simply a feature of the initial conditions. But it would be 
enormously preferable to have a physical theory capable of explaining the 
excess. The necessary ingredients were identified long ago by Sakharov 
\cite{Sak}, and are
\begin{description}
\item[Baryon number violation:] Obviously necessary.
\item[C and CP violation:] {\bf C} is the charge conjugation operator and 
{\bf P} the parity operator. Their violation is necessary in order select a 
preference for baryons or anti-baryons. Their violation is already observed 
in nature in $K$ meson interactions (which do not however violate baryon 
number).
\item[Non-equilibrium conditions:] In equilibrium, reactions occur forwards 
and backwards with the same rate, so even if reactions violate baryon 
number, the forward and backward reactions will cancel out.
\end{description}

The original models for baryogenesis were based on Grand Unified Theories 
for particle interactions, which permit baryons to decay into leptons 
violating baryon number conservation. Non-equilibrium conditions arise 
naturally in an expanding Universe, from the changing relationship between 
the expansion rate and the key particle interaction rates. The standard 
scenario of this kind involves massive particles with baryon number 
violating decays. If their decay is slower than the expansion rate, they are 
unable to stay in equilibrium as the Universe expands, and go through a 
phase of being overabundant before decaying to generate a net baryon number. 
Because the Universe is much cooler by the time they decay, the reverse 
reaction is heavily suppressed. Much work was carried out on this type of 
scenario in the eighties (see Kolb \& Turner \cite{KT} for a review).

In recent years, attention has been focussed in a different direction, 
electro-weak baryogenesis (for a review, see Ref.~\cite{Dol}). It was 
recognized that the electro-weak theory, while preserving baryon number in 
perturbative interactions, could non-perturbatively violate baryon number. A 
configuration doing this has become known as the sphaleron. Although at zero 
temperature this is a tiny effect, it was argued by Kuzmin et al. \cite{KRS} 
that at high temperatures the suppression vanishes. This implies very rapid 
baryon number violation in the early Universe, even if we restrict ourselves 
to Standard Model interactions alone. In fact, strictly speaking it is the 
sum of baryon and lepton numbers which is violated; the difference between 
them, {\bf B}$-${\bf L}, is conserved even by non-perturbative interactions.

Sphalerons have a drastic effect on any pre-existing {\bf B}$+${\bf L} 
symmetry --- they erase it.\footnote{Note that there are effectively no 
observational constraints on the lepton number of the Universe, since the 
neutrino background cannot be directly observed.} Consequently, it appears 
futile to try and make an asymmetry of this type at the GUT era, as it will 
later be destroyed. The simplest GUT, SU(5) [which is in any case ruled out 
by the lack of observed proton decays], can only create an asymmetry of 
this type. However, more complicated GUTs can violate {\bf B}$-${\bf L}, so 
that when {\bf B}$+${\bf L} is driven to zero a residual baryon asymmetry is 
left. Indeed, one interesting proposal is leptogenesis, where a lepton 
asymmetry is generated at high temperatures and the sphalerons used to 
convert part of it into a baryon asymmetry.

Although GUT baryogenesis is clearly then still possible, it would be nice 
to try and capitalize on the baryon number violating property of the 
electro-weak theory to create the baryon asymmetry within the standard 
model. Many proposals have been made to try and realize this, with a belief 
developing that it can only be viable if the electro-weak phase transition 
is quite strongly first-order, so as to maximize departures from 
equilibrium. No compelling model has yet been constructed, with the 
asymmetry coming out usually being too small. However, it is tempting to 
believe that since the answers coming out of these calculations are not 
hopelessly wrong, there must be good chances that these models will be 
shown, after all, to be viable.

\subsection{Topological defects}

Topological defects --- domain walls, cosmic strings, monopoles and textures 
--- offer a rival theory to inflation for the origin of structure (for a 
full account of topological defects, see Vilenkin and Shellard 
\cite{VS}). They are irregularities formed in the Universe during phase 
transitions, and can occur when a scalar field has more than one minimum of 
its potential.

The simplest example is to consider a real scalar field $\phi$ with 
potential
\begin{equation}
V_0(\phi) = \lambda (\phi^2 - M^2)^2
\end{equation}
where $\lambda$ and $M$ are constants. This potential has two minimum, at 
$\phi = \pm M$, and possesses a reflection symmetry $\phi \leftrightarrow 
\phi$.

At high temperatures, this potential is dominated by temperature 
corrections; I'll assume they take a very simple form giving the effective 
potential as
\begin{equation} 
V_T(\phi) = V_0(\phi) + \frac{1}{2} T^2 \phi^2
\end{equation}
At high temperatures, the temperature correction dominates and the minimum 
of the potential is at the origin; this is said to be the symmetric phase. 
However, the Universe cools as it expands and eventually $\phi = 0$ stops 
being the minimum. At low temperature, the field wishes to sit in one of the 
minima, at $\phi = \pm M$.

The lowest energy state of the system is for all the energy to reside in one 
of the minima, everywhere in the Universe. The trouble is, the field has to 
decide in which way to fall, and the existence of horizons means that the 
field in one region of the Universe cannot `signal' to other, causally 
disconnected, regions which direction they are supposed to fall. 
Consequently, in widely separated regions the field makes independent 
choices, and is as likely to fall one way as the other.

The question is, what happens on the boundaries between those regions? The 
field must be continuous, so on the boundary it must smoothly evolve from 
$\phi = - M$ to $\phi = +M$. In doing so it must pass through $\phi = 0$, 
which is a region of high potential energy. Since any line drawn from the 
first point to the second must pass through $\phi = 0$, this potential 
energy must take the form of a sheet, known as a {\bf domain wall}. 

Domain walls are the simplest type of defect that can form, existing where, 
as in the example above, the minimum energy states (known as the {\bf vacuum 
manifold}) are disconnected. More complicated types of vacuum state lead to 
other types of defect; take for example a complex scalar field
\begin{equation}
V = \lambda \left( \phi^{\dagger} \phi - M^2 \right)^2
\end{equation}
Here the minima are at $\phi = M e^{i \alpha}$, where $\alpha \in [0, 
2\pi)$. This vacuum manifold is a circle. There are no domain wall solutions 
(since the vacuum is connected), but it is possible to form a defect known 
as a {\bf cosmic string}.

Cosmic strings form when a loop drawn in space corresponds to a winding 
around the vacuum manifold. The different locations on the loop correspond 
to different angles $\alpha$. Imagine contracting the loop to a point; 
continuity can only occur if the angles become degenerate, which can only 
happen if $\phi = 0$, i.e.~if there is a location within the loop where the 
field is not in the vacuum state. Such locations form a line defect.

More complicated vacuum manifolds lead to more complicated defects; if the 
vacuum is a sphere, then there are no domain walls or cosmic strings, but 
instead a point-like defect known as a magnetic monopole,\footnote{Strictly 
speaking, the theories I've been writing down have what is known as a global 
symmetry, and the defects should be known as global strings and monopoles. 
True cosmic strings and magnetic monopoles occur only when gauge fields 
are included, but the picture of their formation remains as I have 
described here.} which we already encountered as a motivation for inflation. 
An even more exotic type of defect is a texture, corresponding to a 
yet-more-complicated vacuum manifold.

The main cosmological interest in defects is that they are by their very 
nature inhomogeneous. That is, they are able to take a perfectly homogeneous 
Universe before the phase transition and insert inhomogeneities. If the 
defects are massive enough, then they may be seeds for structure formation; 
it turns out that if the symmetry breaking scale is that of Grand Unified 
Theories, then the energy density may be just about right. They therefore 
provide an alternative theory to inflation for the origin of cosmic 
structure.

\begin{figure}[tb]
\centering 
\leavevmode\epsfysize=12cm \epsfbox{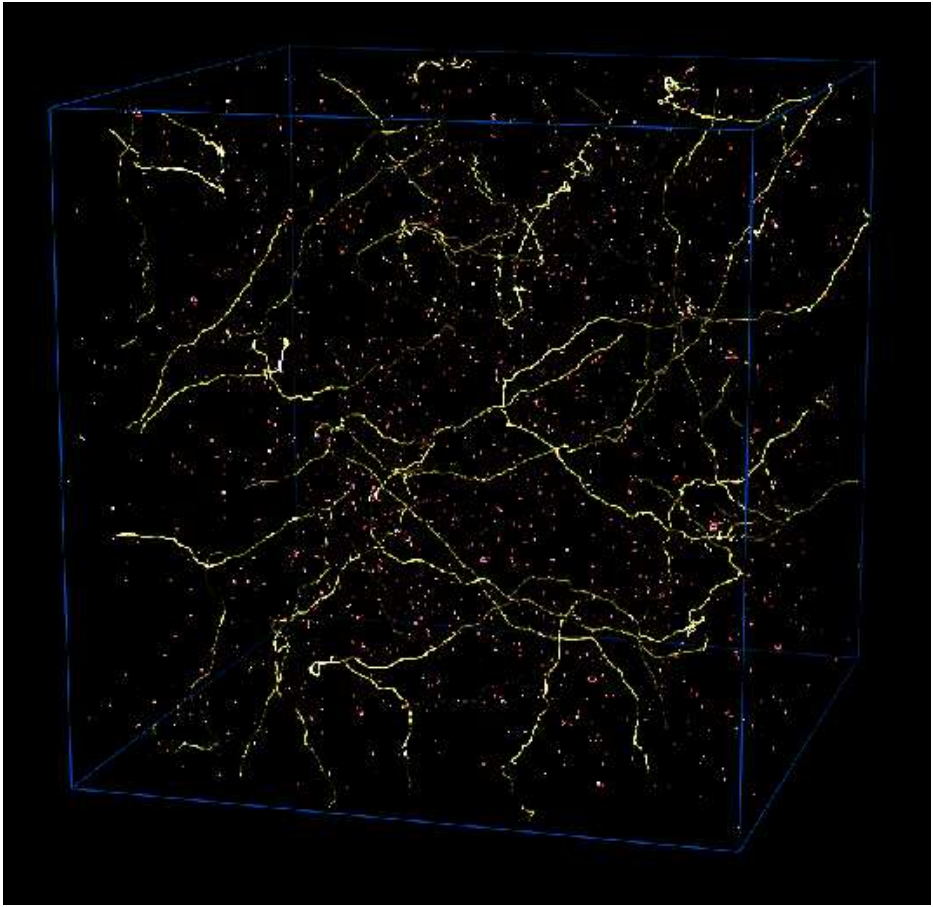}\\ 
\caption[network]{\label{network} A string network can be very complicated; 
this simulation by Allen and Shellard \cite{AS} shows the evolution in the 
matter era. As well as long strings, there are numerous small loops of 
string.} 
\end{figure} 

Unfortunately, the defect theory for structure formation is much more 
complicated than the inflationary paradigm. The reason is because the defect 
dynamics are non-gravitational and non-linear. This means that the 
theoretical status of the subject is some way behind the observational 
situation, whereas inflation is some way ahead. Defects have therefore not 
yet been subjected to the most rigorous tests possible, and indeed may never 
be. Therefore, by far the majority of large-scale structure research has 
been based on the inflationary paradigm, at least implicitly through the 
assumption of gaussian density perturbations evolving only under gravity. I 
imagine that everything else in this conference will fall under the 
inflationary paradigm.

\subsection{Dark matter}

During this School we'll hear a considerable amount of motivation for the 
idea that the bulk of the matter in the Universe is in some as-yet-unknown 
form, known as {\em dark matter}. I won't attempt to review this evidence 
here; instead I'll concentrate on the particle physics aspects.

There are two commonly considered types of dark matter; either the dark 
matter is in the form of elementary particles, or it is in some form of 
compact object. Black holes would appear the most likely candidate of the 
latter type (low mass stars may provide some dark matter but nucleosynthesis 
constraints prevent them from being a candidate for all the dark matter), 
though there is no compelling theory suggesting that they can form in 
sufficient numbers.

Most attention is focussed on the possibility of elementary particles as 
dark matter. There are a range of possibilities, and many experiments are 
now active in the search for particle dark matter.

\subsubsection{Neutrinos}

Within the context of the standard model, there is only one type of particle 
which might possess a significant cosmological density, and that is the 
neutrino. Provided neutrinos are sufficiently light, which turns out to mean 
less than an MeV or so, they are strongly relativistic when they go out of 
thermal equilibrium. This implies that their number density is independent 
of mass (mass being irrelevant in the relativistic limit), and hence their 
energy density is proportional to their mass. A fairly simple calculation 
(see e.g.~Kolb \& Turner) gives their present abundance as
\begin{equation}
\Omega_{\nu} \simeq \frac{\sum_i m_{\nu_i}}{90 h^2 \, {\rm eV}}
\end{equation}
where the sum is over the neutrino families lighter than 1 MeV.  We see that 
a neutrino mass around 30 eV (depending on the precise value of $h$) could 
explain the dark matter, and masses above this are excluded. This number is 
comparable to the experimental bound on the electron neutrino, and well 
below that of the muon or tau neutrino.

Such neutrinos would have been relativistic until fairly recently in the 
history of the Universe, and are known as {\bf hot dark matter}. Evidence 
from structure formation suggests that having all the dark matter as hot 
dark matter will not lead to a satisfactory model, but it remains possible 
for a component of the dark matter to be. Neutrino oscillation experiments 
may be able to probe the relevant mass ranges.

If the neutrinos are more massive than 1 MeV, the relativistic freeze-out no 
longer applies and must be replaced with a non-relativistic one. In this 
regime, the present energy density begins to fall with increasing mass, and 
around a GeV or so we have another solution giving about a critical density 
in neutrinos. Such non-relativistic neutrinos are a form of {\bf cold dark 
matter}.

\subsubsection{WIMPS and axions}

For further dark matter candidates, we must go beyond the Standard Model. 
The most discussed extension is supersymmetry, which at a stroke doubles the 
particle spectrum by associating a boson with every Standard Model fermion 
and a fermion with each Standard Model boson. These so-called superparticles 
possess a new quantum number, {\bf R}-parity, which in the simplest models 
is conserved, guaranteeing that the lightest superparticle is stable. Its 
precise mass depends on the means by which supersymmetry is broken, but 
normally it is some or many GeV and it gives a cold dark matter candidate.

Another alternative is the axion, which arises as a solution to the 
strong-CP problem; that is, it provides a way of suppressing large CP 
violation in strong interactions. The phenomenology of the axion is 
complicated; although it has a very light mass, perhaps $10^{-5}$ eV, it is 
produced non-thermally and gives a cold dark matter candidate.

\subsection{Primordial black holes}

An interesting possibility is that black holes may form at some stage during 
the early Universe. This environment seems to be the only one in which black 
holes might be formed which are light enough that the process of Hawking 
evaporation might be important. Hawking discovered that black holes radiate 
with a temperature given by \cite{Hawk}
\begin{equation}
T_{{\rm BH}} = \frac{\mpl^2}{8\pi M_{{\rm BH}}}
\end{equation}
This gives a lifetime $\tau$ of
\begin{equation}
\tau_{{\rm BH}} = \frac{M_{{\rm BH}}^3}{g_* \mpl^4}
\end{equation}
where $g_*$ is the number of particle degrees of freedom into which the 
black hole can decay. In more readily understood units, this is
\begin{equation}
\frac{\tau_{{\rm BH}}}{10^{10} \, {\rm yrs}} \simeq \left( 
	\frac{M_{{\rm BH}}}{10^{15} \, {\rm grams}} \right)^3
\end{equation}
That is, a black hole with an initial mass of $10^{15}$ grams will have a 
lifetime equal to that of the present age of the Universe, and so will be 
evaporating at the present epoch. Much lighter holes will have evaporated 
long ago, while much heavier ones (such as those formed from stellar 
collapse) will have negligible evaporation.

There are at least three ways in which light primordial black holes can be 
formed during the early Universe.
\begin{enumerate}
\item Black holes may form from large density perturbations 
\cite{BHdens,CGL}, for example induced near the end of an inflationary 
epoch.
\item They may form in phase transitions, particularly strongly first-order 
transitions which proceed explosively by bubble nucleation \cite{BHbub}; 
some inflation models have inflation ending this way.
\item They may form through thermal fluctuations \cite{BHtherm} during the 
very early Universe. This process is only efficient at extremely high 
temperatures, and any holes formed this way would be diluted away were there 
a subsequent period of inflation.
\end{enumerate}

Primordial black holes are observationally interesting, because they 
redshift away as matter (i.e.~as the third power of the scale factor), while 
the early Universe is normally assumed to be radiation dominated. If the 
black holes form early enough, it is therefore quite easy for them to come 
to dominate the energy density of the Universe even if their initial density 
at formation is a tiny fraction of the total. That enables one to place a 
variety of constraints on them.
\begin{itemize}
\item For black holes of masses above $10^{15}$ grams, the only constraint 
is that they must not contribute too much to the energy density of the 
Universe, i.e.~they should not have more than a critical density. If they 
are at this density, then they are a cold dark matter candidate.
\item Black holes evaporating today offer very powerful constraints, most 
importantly those from the $\gamma$ ray background \cite{MacGC}. This limits 
black holes in this mass range to contribute at most orders of magnitude 
below the critical density.
\item Black holes in the range $10^{9}$ grams to $10^{15}$ grams would have 
evaporated at earlier stages in the history of the Universe, and may have 
interfered with early processes such as nucleosynthesis. Again they are very 
strongly constrained.
\item A more speculative possibility is that black holes don't evaporate 
away completely, but instead leave some stable relic. If so, then the 
constraints can strengthen considerably as these relics can contribute 
significantly to the present energy density, even if the initial holes were 
all so light as to have evaporated away \cite{BCL}.
\end{itemize}
A detailed summary of all of these can be found in Carr et al. \cite{CGL}.

\section{Summary}

This has been a brief introduction to a range of Early Universe topics, with 
only inflation covered in any depth at all. The main thrust I have been 
aiming to emphasize is that this area of research is becoming more and more 
a proper area of science, in the sense that models are being falsified and a 
large number of experiments promise to make decisive inroads into 
determining which, if any, of these ideas are on the right track. Progress 
is certain through our improved understanding of structure in the Universe, 
with microwave background satellite experiments poised to reveal the present 
state of the Universe with unprecedented accuracy. More speculatively, 
informative surprises may come our way from a number of sources, for example 
direct detection of dark matter or a convenient nearby supernova.

\section*{Acknowledgments}

The author was supported by the Royal Society.



\section*{Constants and Conversion Factors}

\small
\begin{table}[ht]
\begin{center} 
\begin{tabular}{|lllll|} 
\hline
Newton's constant & $G$~~~~~ & $6.672 \times 10^{-11} \, {\rm m}^3 \, {\rm 
kg}^{-1} \, {\rm sec}^{-2}$ &&\\
Speed of light & $c$ & $2.998 \times 10^{8} \, {\rm m} \, {\rm sec}^{-1}$ 
& or & $3.076 \times 10^{-7} \, {\rm Mpc} \, {\rm yr}^{-1}$\\
Reduced Planck constant & $\hbar= h/2\pi$ & $1.055 \times 10^{-34} \, {\rm 
m}^2 \, {\rm kg} \, {\rm sec}^{-1}$ &&\\
Boltzmann constant & $k_{{\rm B}}$ & $ 1.381 \times 10^{-23} \,{\rm J} \, 
{\rm K}^{-1}$ 
& or & $8.619 \times 10^{-5} \, {\rm eV} \, {\rm K}^{-1}$ \\
Radiation constant & $\alpha$ & $7.565 \times 10^{-16} \, {\rm J} \, {\rm 
m}^{-3} \, {\rm K}^{-4}$ && \\
Planck mass & $m_{{\rm Pl}}$ & $2.179 \times 10^{-8} \, {\rm kg}$ & or &
$1.22 \times 10^{19} \, {\rm GeV}$\\
\hline
\end{tabular} 
\end{center} 
\caption[constants]{Some fundamental constants.}
\end{table} 

\begin{table}[ht]
\begin{center} 
\begin{tabular}{|c|}
\hline
$1 \, {\rm pc} = 3.262$ light years $= 3.086 \times 10^{16} \, {\rm m}$\\
$1 \, {\rm yr} = 3.16 \times 10^7 \, {\rm sec}$ \\
$1 \, {\rm eV} = 1.602 \times 10^{-19} \, {\rm J}$\\
$1 \, \msun = 1.989 \times 10^{30} \, {\rm kg}$\\
$1 \, {\rm J} = 1 \, {\rm kg} \, {\rm m}^2 \, {\rm sec}^{-2}$\\
\hline
\end{tabular} 
\end{center} 
\caption[conversion]{Some conversion factors.}
\end{table} 

\end{document}